# Ultrafast Photodynamics of Glucose

Jens Petersen,* Joachim O. Lindner, and Roland Mitrić*

*Julius-Maximilians-Universität Würzburg, Institut für physikalische und theoretische Chemie, Emil-Fischer-Str. 42, 97074 Würzburg, Germany*

E-mail: jens.petersen@uni-wuerzburg.de; roland.mitric@uni-wuerzburg.de

Phone: 0049 931 31 88832; 0049 931 31 85135

**Abstract**

We have investigated the photodynamics of $\beta$-D-glucose employing our field-induced surface hopping method (FISH), which allows us to simulate the coupled electron-nuclear dynamics, including explicitly nonadiabatic effects and light-induced excitation. Our results reveal that from the initially populated $S_1$ and $S_2$ states, glucose returns nonradiatively to the ground state within about 200 fs. This takes place mainly via conical intersections (CIs) whose geometries in most cases involve the elongation of a single O-H bond, while in some instances ring-opening due to dissociation of a C-O bond is observed. Experimentally, excitation to a distinct excited electronic state is improbable due to the presence of a dense manifold of states bearing similar oscillator strengths. Our FISH simulations explicitly including a UV laser pulse of 6.43 eV photon energy reveals that after initial excitation the population is almost equally spread over several close-lying electronic states. This is followed by a fast nonradiative decay on the time scale of 100-200 fs, with the final return to the ground state proceeding via the $S_1$ state through the same types of CIs as observed in the field-free simulations.

# 1 Introduction

Carbohydrates represent one of the most abundant classes of molecules present in biological systems. Owing to their extraordinarily rich structural variability, they are capable of numerous biological functions, e.g. as parts of nucleic acids for encoding genetic information or of membrane glycoproteins for molecular recognition and thus for signalling processes in living cells. Moreover, they are utilized by Nature as an energy storage, e.g. in polymers like starch, as well as for building cellular structures, most prominently in the form of cellulose, which is a crucial constituent of the cell walls of all green plants.

While there has been a large and long-standing effort in determining structural and chemical properties of carbohydrates both experimentally and theoretically (cf.[1–3] and references therein), only little is known about their photochemistry and photophysics. This is mainly due to the lack of chromophoric units in these molecules, for which reason instead of sharp, energetically low-lying absorption features only a broad absorption in the vacuum UV region is observed. Therefore, in natural terrestrial environment, photodynamical processes in carbohydrates play a minor role, and direct spectroscopical investigations are, due to the large excitation energies, impeded as well. However, the situation is strongly different under extraterrestrial conditions, where high-energetic radiation is abundant. Indeed, a recent study showed that under conditions mimicking interstellar ice grains bio-organic molecules, amongst them carbohydrates up to ribose, are formed from simple precursors such as water, methanol and ammonia.[4] In this environment, photochemical reactions are likely to take place, and especially the question of photostability of the formed organic molecules is highly relevant in the context of chemical evolution and the origins of life.

What is experimentally known since quite a long time is that UV irradiation of aqueous carbohydrate solutions applying wavelengths down to 200 nm invokes photochemical reactions resulting in the formation of products which exhibit intense absorption between 240 and 270 nm[5,6]. In studies of the glucose molecule as a prototype example, it has been found that this is due to photodegradation processes in which molecules containing chromophoric subunits are formed, with the degradation process starting at the anomeric $C^1$ atom of the carbon chain[7]. This has lead to the suggestion that the electron lone pairs of the neighbouring oxygen atom in the pyranose ring may be involved in the photoexcited electronic states, giving rise to their $n\sigma^*$ character[7]. Such type of electronic states, alongside with those of $\pi\sigma^*$ character, have been intensively investigated theoretically in the recent years in a variety of organic molecules, and their general dissociative nature has been established[8–13]. However, detailed theoretical studies of the photophysical and photochemical properties of carbohydrates have been lacking until very recently[14,15].

On the experimental side, the fact that continuous UV irradiation of carbohydrates leads to strongly UV-absorbing photoproducts has been exploited in the last years in the context of analytical techniques, such as the quantification of carbohydrate contents in beverages by spectrophotometric detection[16], or as an alternative detection approach in capillary electrophoresis[17–19]. Usually, the detection process within the latter method involves UV absorption of the separated analytes. If the analyte molecule lacks a chromophoric subunit, as in the case of carbohydrates, then, frequently, chemical modification is performed, e.g. a chromophoric moiety is inserted, thus allowing for facile UV detection. An alternative approach, however, is based on irradiating the analytes with UV light of wavelengths down to 180 nm. This leads to photodegradation and the subsequent formation of reaction products that exhibit UV absorption above 260 nm, which is easily accessible to the standard detection techniques[17,18]. Investigation of the reaction products formed from glucose samples in this way has also confirmed that the degradation process is initiated by a ring opening reaction in which the bond between the $C^1$ carbon and the ring oxygen is broken.[18,19]

A detailed theoretical insight into the optical absorption and possible photochemical relaxation pathways of the glucose molecule has been gained by investigations of Tuna et al.[14]. Glucose represents a building block found in various naturally occuring polymeric carbohydrates and thus can be considered as a basic representative of this class of molecules. Performing quantum chemical calculations of electronically excited states as well as determining the locations of conical intersections between ground and first excited electronic state, Tuna et al. identified possible channels by which electronically excited glucose may relax back to the ground state. Specifically, two main classes of conical intersections were found, in which either an O-H bond is elongated or the ring system of glucose is opened at the anomeric carbon atom. By assuming such deformed structures, the molecules may efficiently relax to the ground state, followed either by additional dissociative processes or by restoring the original molecules - however, with increased internal vibrational energy. Very similar findings have meanwhile also been obtained for the ribose molecule[15].

In the present contribution, we aim to complement the aforementioned static calculations by an investigation of the dynamical processes occurring in photoexcited glucose. We will address both the specific relaxation mechanism following the population of particular excited states, as well as the photophysics induced by interaction of the molecule with a laser pulse of nonzero spectral width. For this purpose, nonadiabatic molecular dynamics simulations in the framework of our field-induced surface hopping method (FISH)[20] have been carried out, and the time scales of nonradiative relaxation as well as the nature of the corresponding nuclear rearrangements have been analysed.

## 2 Computational methods

The structure of $\beta$-D-glucose was optimized and the harmonic frequencies were determined using the Turbomole program package[21] in the frame of three different approaches: density functional theory (DFT) employing the B3LYP functional[22] and (i) the 6-31++G basis set[23,24] as well as (ii) the 6-31++G** basis set[23–25] and (iii) second order Møller-Plesset perturbation theory (MP2) employing the aug-cc-pVDZ basis set[26,27]. Subsequently, the normal modes were used to sample a Wigner distribution corresponding to a temperature of 250 K for each method, from which initial coordinates and momenta for use in the dynamics simulations were generated. The nonradiative relaxation dynamics of electronically excited glucose was simulated in the frame of Tully's surface hopping method[28]. For this purpose, the electronic energies and gradients were obtained "on the fly" using Turbomole with the same DFT functional and basis set as stated above along classically propagated nuclear trajectories. The excited state energies and gradients were calculated using linear-response time-dependent density functional theory (LR-TDDFT)[29]. The nonadiabatic coupling elements were calculated from the LR-TDDFT eigenvectors according to the procedure of Mitrić et al.[30]. Using these quantities, the electronic population dynamics was simulated along each trajectory by solving the time-dependent Schrödinger equation, and the resulting populations were employed to compute hopping probabilities according to Ref.[31,32]. In order to ensure conservation of the total energy of each trajectory, the velocities were uniformly rescaled after a successful surface hop. The propagated classical ensemble consisted of 30 trajectories which were initiated in the $S_1$ and $S_2$ state, respectively. The propagation starting from the $S_1$ state was performed using both the 6-31++G basis set and its counterpart containing polarization functions, 6-31++G**. The results using these two basis sets provide essentially identical excited state lifetimes and reaction channels (cf. Supporting Information, Fig. S1-S3). In particular, the process of O-H bond stretching, which is important in the excited state dynamics of glucose (cf. Results and Discussion) is equally well described by the two approaches, as can be also inferred from the very similar behaviour of the $S_0$ and $S_1$ electronic state energies along the O-H bond stretching coordinate, which is presented in Fig. S4 of the Supporting Information. Therefore, the further simulations were performed using the computationally more efficient basis set 6-31++G. For the propagation of the nuclei, the classical Newton's equations were solved numerically using the velocity Verlet algorithm with a time step of 0.25 fs. The total propagation time was 200 fs. When reaching regions near $S_0$-$S_1$ conical intersections, which could not be accurately described by TD-DFT[33], those trajectories which were previously situated in the $S_1$ state were continued in the ground state, while those residing in higher electronic states were aborted. Trajectories were subsequently classified according to their structure at the conical intersections.

In order to provide a comparison of the LR-TDDFT approach for the nonadiabatic dynamics to an ab initio-based methodology, we performed additional surface-hopping simulations using an ensemble of 30 trajectories starting from the $S_1$ state and employing the algebraic diagrammatic construction through second order (ADC(2)) method[34] together with the aug-cc-pVDZ basis set[26,27]. The implementation of the nonadiabatic couplings straightforwardly follows our approach for LR-TDDFT[30], which has been also previously employed by Plasser et al.[35].

For a more realistic description of the laser excitation into a manifold of close-lying states with comparable oscillator strengths, in addition to the field-free nonadiabatic dynamics simulations also the field-induced surface hopping method (FISH)[20] was applied, again employing the TDDFT (B3LYP/6-31++G) approach. For this purpose, 100 trajectories sampled from the Wigner function described above were propagated over 300 fs within a manifold of 10 electronic states. The excitation field was taken to be of Gaussian shape, with a field intensity of $7.9 \cdot 10^{12} \frac{\text{W}}{\text{cm}^2}$, a central frequency corresponding to 6.43 eV and a full width at half maximum (FWHM) of 30 fs. Unlike the field-free simulations, the total energy of a given trajectory is not conserved while the laser pulse is present. Therefore, velocity rescaling after a surface hopping event was only enabled for the time period after 50 fs, when the pulse had ceased.

# 3 Results and Discussion

## 3.1 UV absorption spectrum

The calculated absorption spectrum of $\beta$-glucose, obtained from the 20 lowest excited electronic states in the frame of TDDFT (B3LYP/6-31++G) for each of the 100 initial conditions sampled from a harmonic Wigner distribution at 250 K, is presented in Fig. 1. It is characterized by a weak onset around 5.2 eV, with appreciable oscillator strengths above 5.5 eV. Up to 7 eV, the oscillator strength steadily increases. Its subsequent drop is due to the limited number of states included in the calculation. Compared to experimental data obtained in aqueous solution[7], the absorption onset is red-shifted by about 0.5 eV. As has been shown by Tuna et al.[14], the energetic position of the excited states is slightly underestimated by TDDFT, and using correlated ab initio methods such as CC2 leads to a blue shift of ∼0.3 eV. The electronic state character was shown to be essentially the same in both methods. This gave us confidence to proceed with employing the computationally less demanding TDDFT approach for dynamics simulations.

The electronic absorption of glucose is characterized by the absence of distinct intense excited states. Instead, the energy range up to 7 eV contains already about 10 excited states with relatively low oscillator strengths. For the optimized ground state structure, which is also shown in Fig. 1, the analysis of the lowest 10 excited states is summarized in Table 1, and the relevant molecular orbitals are depicted in Fig. 2. The excited states under investigation all share similar electronic character, with the leading excitations occuring from orbitals representing linear combinations of oxygen lone pairs to Rydberg-type virtual orbitals. It has been shown previously that these Rydberg orbitals change their nature upon elongation of C-O or O-H bonds, ultimately assuming $\sigma^*$ character[14]. Therefore, the typical dissociative behaviour of n$\sigma^*$ states is expected to show up in the photodynamics.

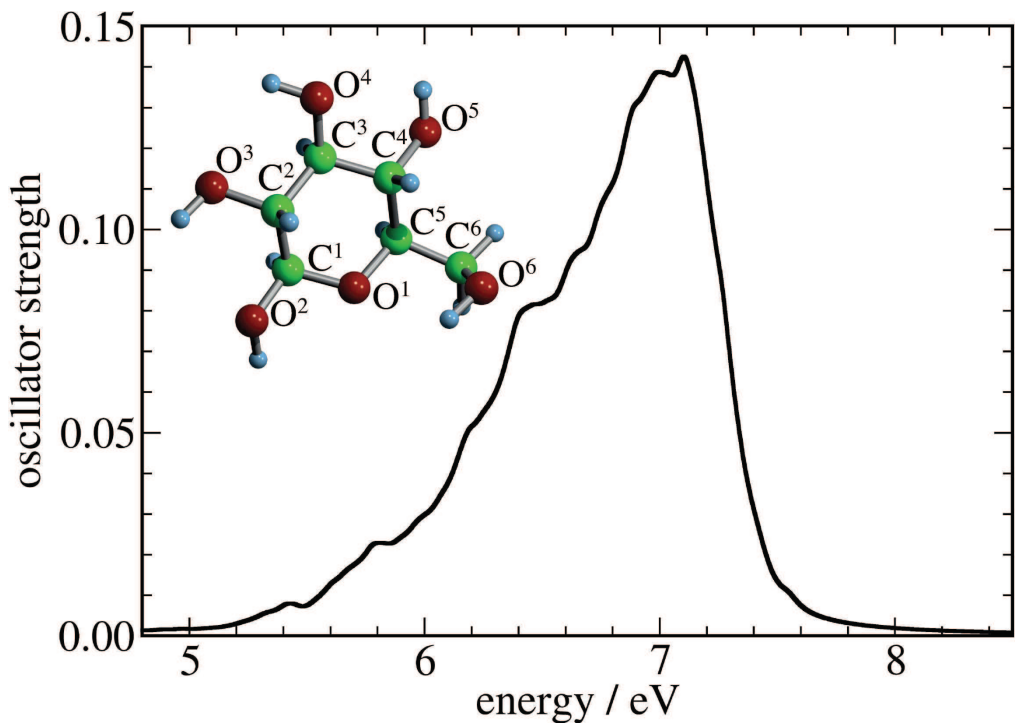

Figure 1: Electronic absorption spectrum of the ensemble of 100 initial conditions. For each structure, the lowest-lying 20 excited states have been calculated using TD-DFT (B3LYP/6-31++G). The individual transitions have been convolved by a Lorentzian width of 0.1 eV. The optimized structure of $\beta$-D-glucose is shown as an inset.

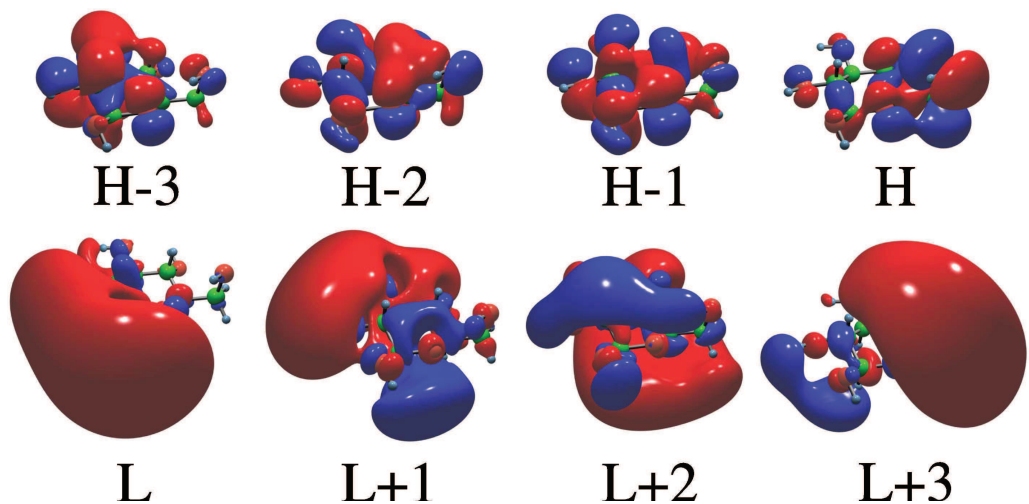

Figure 2: Molecular Kohn-Sham orbitals for the optimized structure of $\beta$-D-glucose. H denotes the highest occupied, L the lowest unoccupied molecular orbital.

Table 1: Excited electronic states for the optimized structure of $\beta$-D-glucose obtained by TDDFT (B3LYP, 6-31++G).

| State | Transition energy (eV) | Oscillator strength | Composition (%) |
|---|---|---|---|
| $S_1$ | 6.284 | 0.0033 | H→L (48) |
|  |  |  | H−1→L (39) |
| $S_2$ | 6.410 | 0.0042 | H−1→L (37) |
|  |  |  | H→L (31) |
|  |  |  | H→L+1 (12) |
| $S_3$ | 6.543 | 0.0004 | H−1→L+1 (59) |
|  |  |  | H→L+1 (19) |
| $S_4$ | 6.602 | 0.0155 | H→L+1 (34) |
|  |  |  | H−1→L+1 (25) |
| $S_5$ | 6.680 | 0.0003 | H−2→L+1 (35) |
|  |  |  | H−2→L (16) |
|  |  |  | H→L+3 (14) |
| $S_6$ | 6.713 | 0.0051 | H−2→L (42) |
|  |  |  | H→L+3 (14) |
| $S_7$ | 6.757 | 0.0074 | H−2→L (22) |
|  |  |  | H−3→L+1 (20) |
|  |  |  | H−3→L (17) |
|  |  |  | H→L+3 (10) |
| $S_8$ | 6.816 | 0.0015 | H→L+2 (25) |
|  |  |  | H→L+1 (19) |
|  |  |  | H−2→L+1 (12) |
| $S_9$ | 6.886 | 0.0027 | H→L+2 (51) |
|  |  |  | H→L+3 (25) |
| $S_{10}$ | 6.954 | 0.0075 | H−1→L+2 (65) |
|  |  |  | H−3→L (17) |

## 3.2 Field-free surface-hopping dynamics

For the investigation of dynamical processes occurring in glucose after photoexcitation, in a first step, the nonradiative relaxation of the molecule has been simulated starting from the first or second excited electronic state, respectively. The electronic state population dynamics for an ensemble of 30 trajectories initiated in the $S_1$ state is shown in Fig. 3. It is evident that, besides transient population of the $S_2$ state, fast internal conversion to the ground state takes place, which is completed after 200 fs. The mechanism of this ultrafast process can be unravelled by analysing some typical trajectories. In Fig. 4, a short trajectory is presented in which the dynamics in the $S_1$ state straightforwardly leads to a conical intersection (CI) with the ground state which is reached already after 15 fs. This is accompanied by elongation of the $O^2$-H bond (cf. structure shown in Fig. 1). During the dynamics, the excited $S_1$ state is of $n\sigma^*$ character, which is consistent with previous findings of the dissociative nature of such states[14].

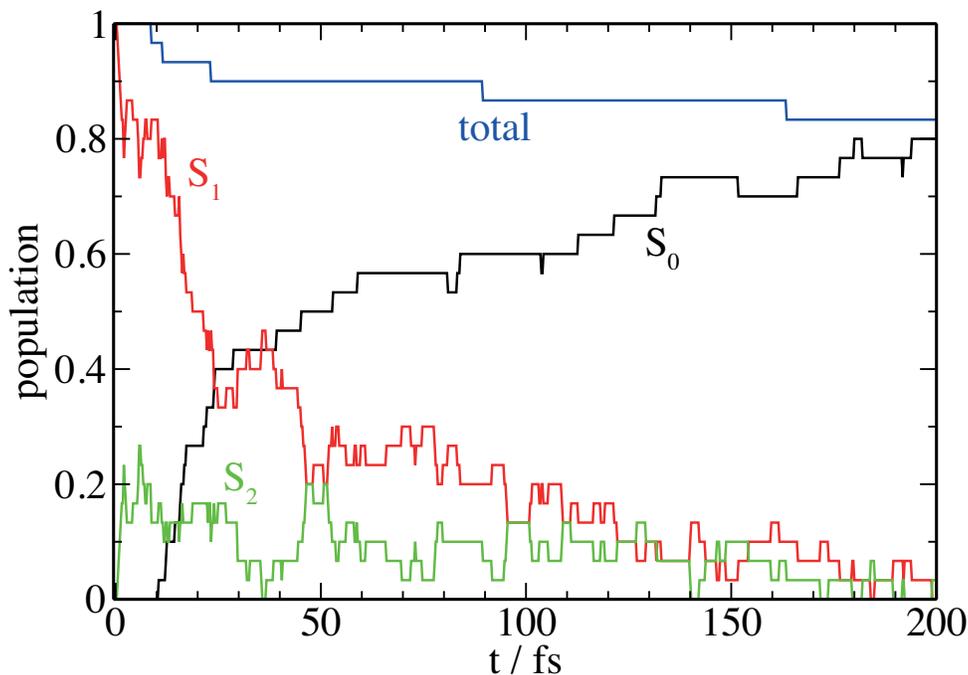

Figure 3: Time-dependent electronic state populations for nonadiabatic dynamics starting in the $S_1$ state. The decrease of the total population is due to such trajectories that resided in the $S_2$ state when reaching the vicinity of an $S_0$-$S_1$ CI, where the propagation was stopped.

Besides O-H bond elongation, opening of the pyranose ring has been observed as a second photochemical channel, as illustrated in Fig. 5. In this case, the initial stage of the dynamics is similar to the one shown in Fig. 4, leading first to elongation of an O-H bond and closing of the $S_1$-$S_0$ energy gap. However, before reaching the conical intersection, a hop to the ground state occurs. This leads to energy transfer from the electronic degrees of freedom to the nuclear vibrations, especially invoking large-amplitude vibrations of the O-H bond. Nonetheless, bond breaking does not take place in the ground state here. Rather, the large vibrational energy together with the fact that the electronic system resides in a superposition of the $S_0$ and $S_1$ states enables the trajectory to hop back to the $S_1$ state at 80 fs. Subsequently, the motion is directed towards a conical intersection characterized by a ring-opening motion (cf. the increased $C^1$-$O^1$ distance shown in the lower part of Fig. 5).

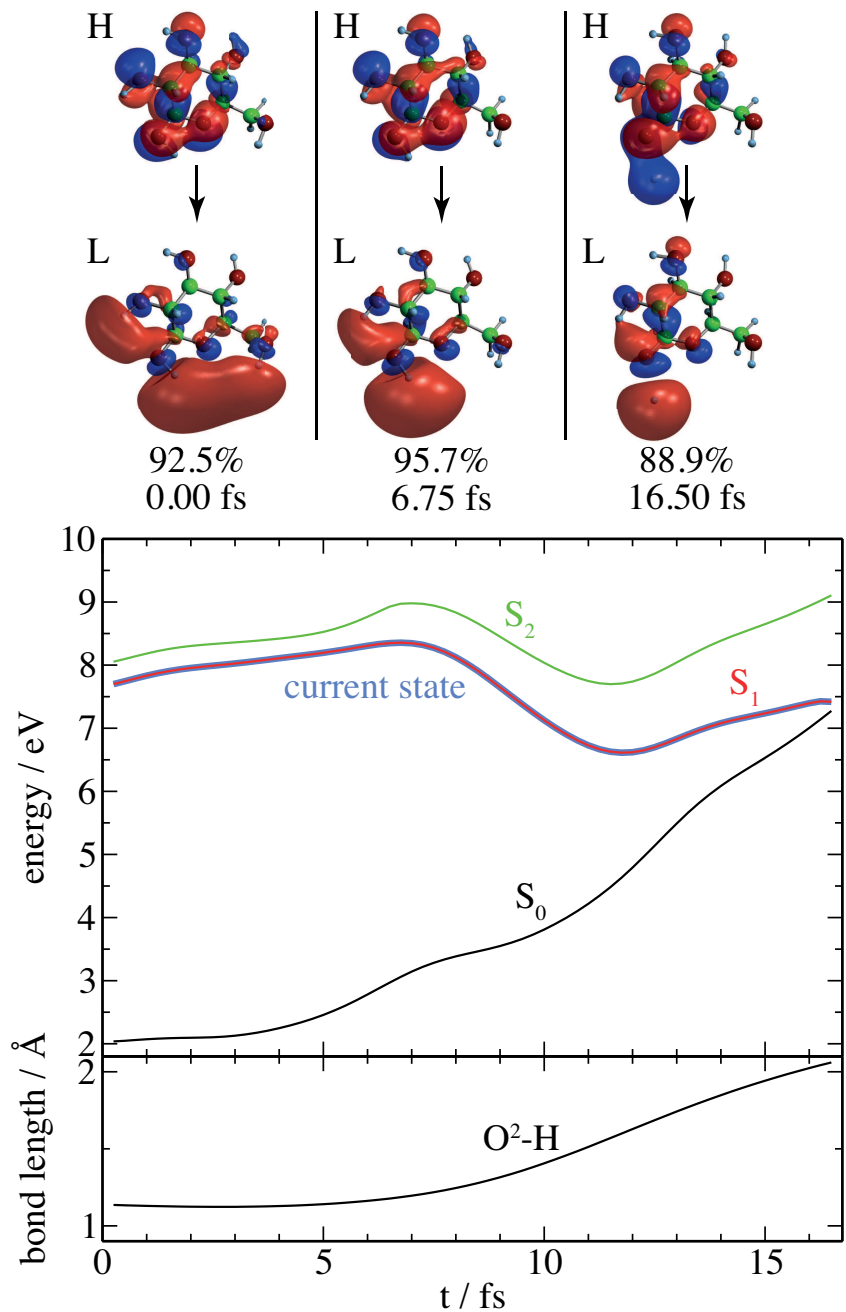

Figure 4: (Middle) Energies of the lowest 3 electronic states along a selected trajectory initiated in the $S_1$ state, where the $O^2$-H-elongation leads to a conical intersection. Energies are given relative to the optimized geometry of the ground state. (Top) Molecular orbitals showing the main excitation at distinct points in time. (Bottom) $O^2$-H bond length along the trajectory.

These examples illustrate the two general possibilities found for nonradiative decay in the glucose molecule: Either the return to the ground state proceeds via a CI connected to O-H bond elongation, or via another one which invokes opening of the pyranose ring. Which path a given trajectory actually follows depends sensitively on the initial conditions. As a general trend, the CIs connected to O-H bond elongation are reached much more rapidly, as can be inferred from Fig. 6, in which the energies and arrival times of all conical intersections traversed by the propagated trajectories are presented. These O-H-dissociative CIs are mostly reached within 30 fs of propagation time, whereas to arrive at the CIs connected to ring opening, usually a time between 100 and 200 fs is needed. It is interesting to note that the OH-dissociative CIs tend to lie energetically above the ring-opening ones, but are still reached faster and by the majority of trajectories (about 67 % vs. 23 %). This confirms previous suggestions of Tuna et al. who investigated static reaction paths connecting the Franck-Condon region with the two types of conical intersections discussed above and predicted the O-H dissociative CIs to be dominantly involved in the dynamical relaxation processes[14]. A possible explanation for this finding can be based on the actual distances along the pathways the trajectories have to cover in the 3N dimensional configuration space from the initial points of the dynamics to the vicinity of the respective CI. For the two subsets of trajectories reaching either the O-H dissociative CIs or the C-O ring opening ones, these distances amount to 7.8 vs. 24.4 Å in average, clearly indicating that the O-H CIs are reached along a shorter path. Although for most of the trajectories the return to the ground state involves the passage through CIs in which specific bonds are elongated, the actual breaking of bonds in the ensuing ground state dynamics is only observed as a minor channel, about 80 % of the trajectories finally returning to the original cyclic form of the molecule. In the remaining cases, reactions such as abstraction of H or $H_2O$ or forming of open-chain molecules are observed.

In order to provide a comparison between our TDDFT-based approach for the nonadiabatic dynamics and a fully ab-initio-based methodology, we have performed additional surface-hopping simulations starting in the $S_1$ state in the frame of the ADC(2) method as described in the Computational Section. The resulting population dynamics, presented in Fig. 7, shows that the general time scale of the nonadiabatic return to the ground state is about 100 fs, slightly faster than in the case of TDDFT, but of the same order of magnitude. Similarly, also the pathways taken towards the conical intersections clearly correspond to those seen in the TDDFT simulations. The majority of trajectories approaches the $S_0/S_1$ degeneracy via O-H bond stretching in a sub-50 fs time scale. An illustrative trajectory for this process is presented in Fig. 8, where the CI is reached already after 9 fs. A much smaller part of the trajectories follows a path where the pyranose ring of the molecule opens when the CI is approached, as is exemplified by the trajectory shown in Fig. 9. Also in agreement with the TDDFT simulations, this process takes longer time, more than 50 fs in the present case. Regarding the fate of the trajectories after return to the ground state, the ADC(2) dynamics, which reduces to MP2 in the ground state, shows a high amount of completed O-H bond dissociations, which have only rarely been observed in the TDDFT simulations. Closer inspection of the electronic energies along an O-H bond stretching coordinate, as shown in Fig. S4 of the Supporting Information, reveals that this is due to a continuous decrease of the electronic energies for increasing bond lengths, while in the case of TDDFT an asymptotically constant value is adopted. Due to a well-known deficiency of MP2 in describing situations where degenerate occupied and virtual orbitals occur, such as homolytic bond breaking,[36] this behaviour of the trajectories can be clearly qualified as incorrect, and the MP2 ground state dynamics following the passage through the CI is not further taken into consideration.

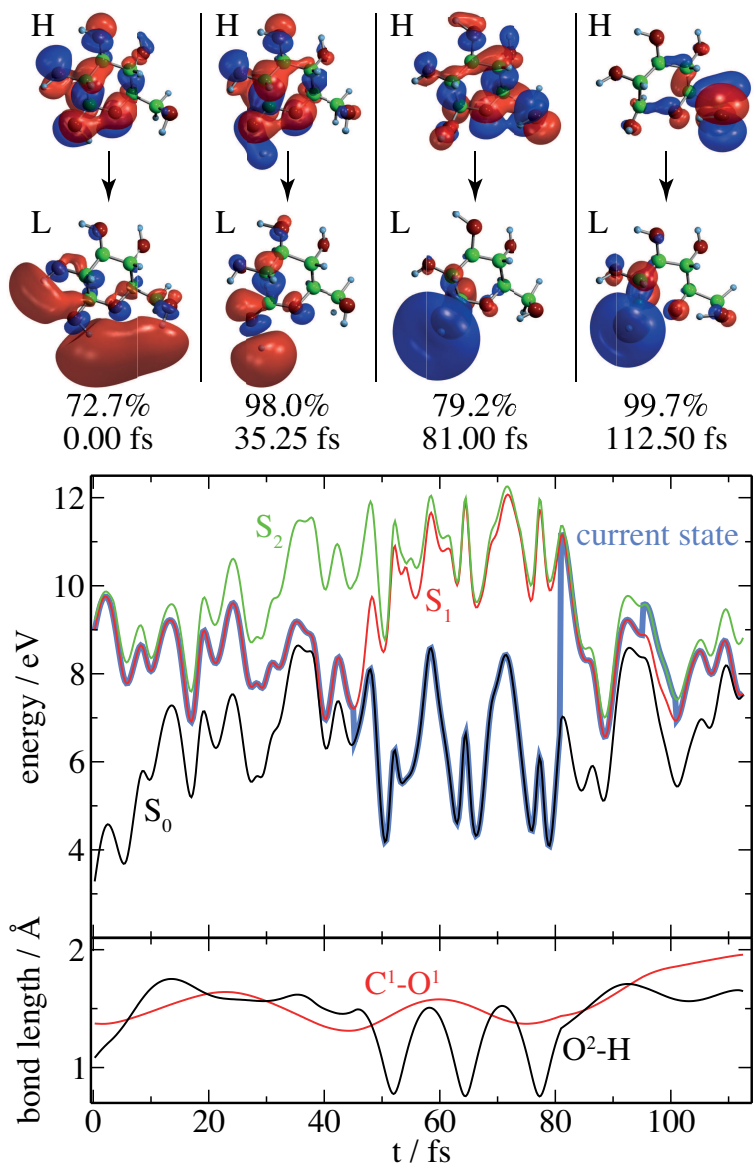

Figure 5: (Middle) Energies of the lowest 3 electronic states along a selected trajectory initiated in the $S_1$ state, where the $C^1$-$O^1$ ring-opening leads to a conical intersection. Energies are given relative to the optimized geometry of the ground state. (Top) Molecular orbitals showing the main excitation at distinct points in time. (Bottom) $O^2$-H and $C^1$-$O^1$ bond lengths along the trajectory.

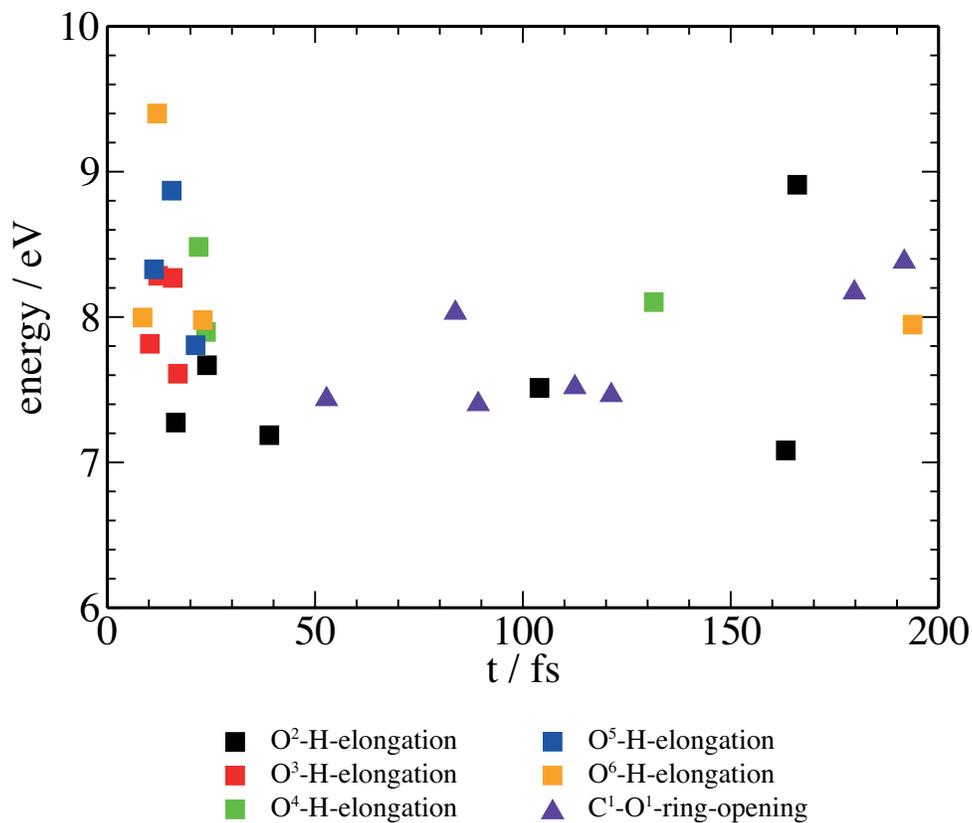

Figure 6: Ground state energies of the trajectories initiated in the $S_1$ state at the final time steps when the TDDFT energy gap between the $S_0$ and $S_1$ states closes. Energies are given relative to the optimized geometry of the ground state. The different types of conical intersections (CIs) reached are classified according to the respective geometry changes as O-H elongation (squares) and as ring-opening (triangles). The CIs characterized by O-H elongation are further colour-coded according to which specific bond is affected.

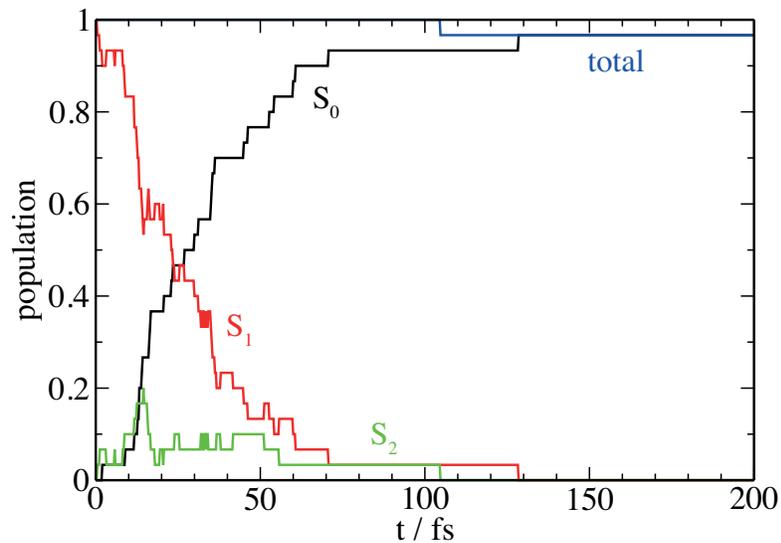

Figure 7: Time-dependent electronic state populations for nonadiabatic ADC(2) dynamics starting in the $S_1$ state.

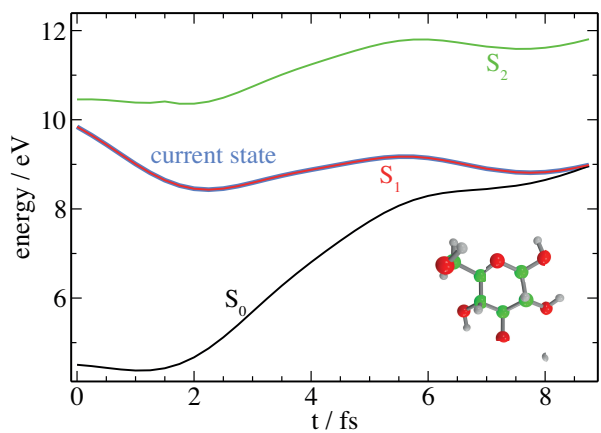

Figure 8: Energies of the lowest 3 electronic states along a selected ADC(2) trajectory initiated in the $S_1$ state, where the O-H-elongation leads to a conical intersection. Energies are given relative to the optimized geometry of the ground state.

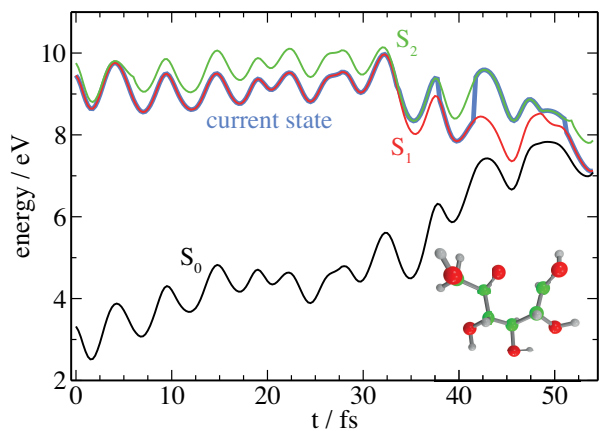

Figure 9: Energies of the lowest 3 electronic states along a selected ADC(2) trajectory initiated in the $S_1$ state, where the $C^1$-$O^1$ ring-opening leads to a conical intersection. Energies are given relative to the optimized geometry of the ground state.

Since the electronic absorption spectrum of glucose is characterized by many excited states of similar oscillator strength, the investigation of nonadiabatic relaxation cannot be restricted to simulations starting in the lowest excited state only. Therefore, we studied the influence of the initial state selection on the course of the dynamics by performing additional simulations started in the $S_2$ state. The resulting population dynamics is presented in Fig. 10 and shows that from the initial $S_2$ state the population transfer takes place rapidly (within less than 20 fs) to the $S_1$ state, which reaches a peak population of 60 % after 10 fs. Subsequently, the population of both states decays on a time scale of 100 - 150 fs. After the simulation time of 200 fs, the population has essentially completely returned to the ground state.

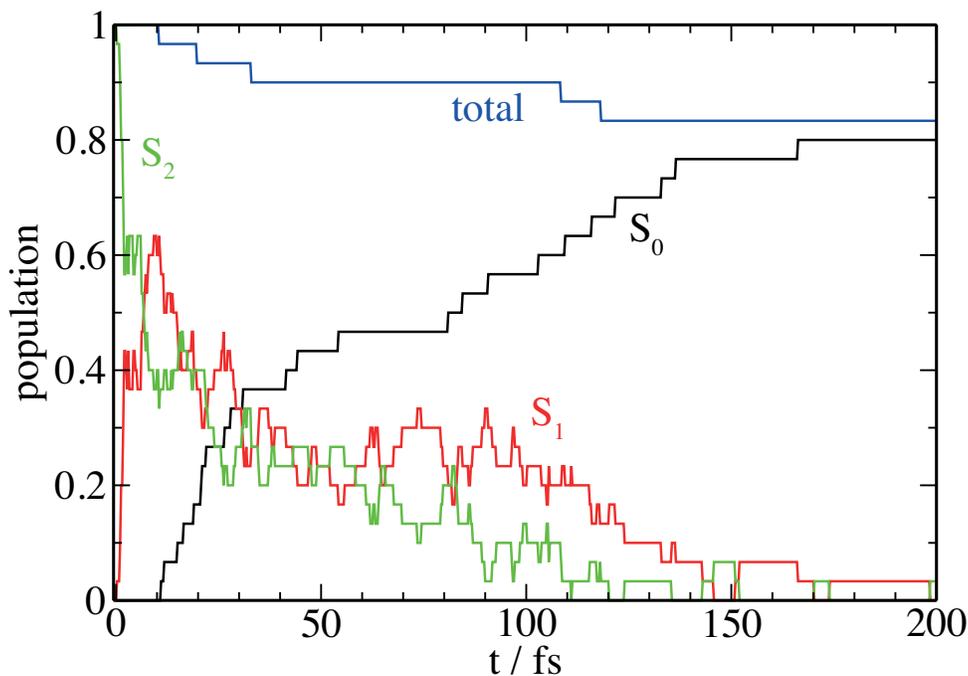

Figure 10: Time-dependent electronic state populations for nonadiabatic dynamics starting in the $S_2$ state. The decrease of the total population is due to such trajectories that resided in the $S_2$ state when reaching the vicinity of an $S_0$-$S_1$ CI, where the propagation was stopped.

The decay mechanisms found in this simulation are very similar to those occurring for trajectories started in the $S_1$ state. In particular, this is the case for trajectories that approach an O-H-dissociative CI via an $n\sigma^*$ excited state, as shown in Fig. 11. Although started in the $S_2$ state, hopping to the $S_1$ state takes place very fast and the approach to the CI again occurs on a short time scale of about 20 fs.

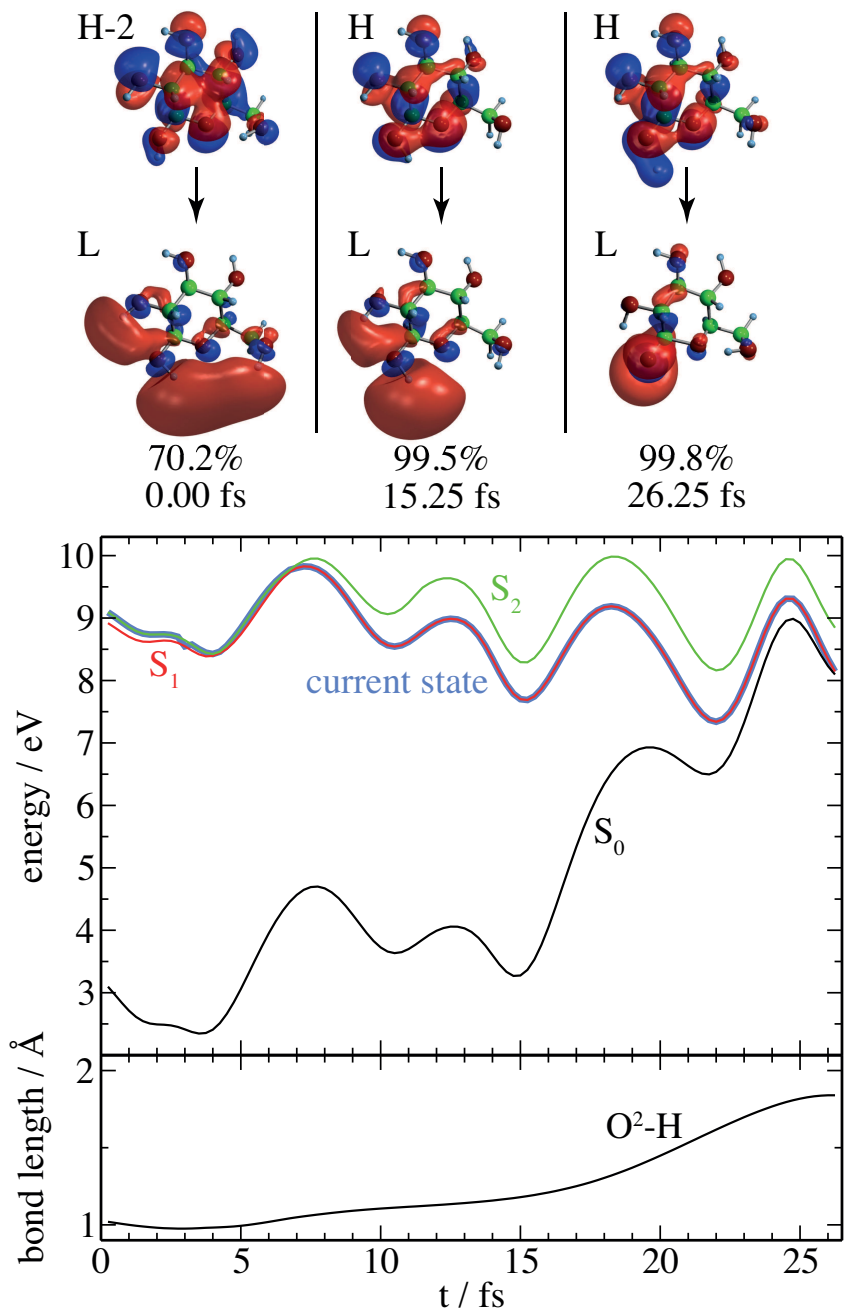

Figure 11: (Middle) Energies of the lowest 3 electronic states along a selected trajectory initiated in the $S_2$ state, where the $O^2$-H-elongation leads to a conical intersection. Energies are given relative to the optimized geometry of the ground state. (Top) Molecular orbitals showing the main excitation at distinct points in time. (Bottom) $O^2$-H bond length along the trajectory.

Besides O-H dissociation, deactivation via ring opening is observed as well. In the case of trajectories initiated in the $S_1$ state, it has already become evident that in order to reach a ring-opening CI, the molecule has to bypass the O-H dissociative CIs which lie much closer to the starting geometry of the dynamics. In the example presented in Fig. 5 above, this was possible since the trajectory intermediately changed to the ground state, and only after some time, its large vibrational energy allowed it to populate the excited state again. The illustrative trajectory shown in Fig. 12 behaves differently. Here, the regions of the potential energy surface which are close to the O-H dissociative CI, characterized by elongation of the O-H bond (cf. lower part of Fig. 12), are avoided by the fact that the molecule still resides in the $S_2$ state when approaching geometries of elongated O-H bonds. Therefore, no direct transition to the ground state is possible. Instead, the dynamics proceeds for some time on the $S_2$ state before switching to the $S_1$ state after 45 fs. Now the molecule is situated in a region of the $S_1$ potential energy surface from where access to a ring-opening CI is possible, which is reached after a propagation time of about 100 fs.

The increased possibilities to avoid the direct approach to O-H dissociative CIs are also reflected by the fact that a higher number of trajectories undergoes ring-opening after a propagation time of more than 70 fs, as can be inferred from Fig. 13. However, the main deactivation channel remains O-H dissociation on a sub-50-fs time scale. As for the trajectories initiated in the $S_1$ state, the lengths of the pathways between the initial geometry and the vicinity of the CI have been calculated, resulting in similar average values of 8.0 and 19.8 Å for the O-H dissociative and C-O ring-opening CIs, respectively. Therefore, also in this case the pathways towards the O-H dissociative CIs are distinctly shorter. After return to the ground state, the majority of the molecules re-assumes the cyclic form of glucose without undergoing a chemical reaction, also similar to the dynamics initiated in the $S_1$ state.

### 3.3 Field-induced surface hopping dynamics

In order to achieve a more realistic simulation of the initial light-induced excitation processes in glucose, additional FISH simulations with explicit light excitation have been carried out, in which a pump laser pulse of 193 nm (6.43 eV) central wavelength and a temporal FWHM of 30 fs was employed. Given the high number of energetically close-lying excited states, which also bear comparable oscillator strengths, the light pulse populates a manifold of electronic states rather than a single one, as can be seen from the time-dependent populations presented in Fig. 14. The overall excitation efficiency is about 80 %. However, 25 % of the trajectories already return to the ground state within the laser pulse duration due to the coupling with the field. After the field has ceased, the photoexcited trajectories exhibit several mechanisms for nonradiative decay. In general, the deactivation is fast and essentially completed after 300 fs, regardless of the specific states populated during the laser excitation. The transition to the ground state proceeds through the same types of CIs as found in the field-free dynamics. However, due to the intake of energy in the molecular system as a consequence of the light-molecule interaction, the temporal and energetic distribution of the CIs reached during the dynamics is different, as can be seen from Fig. 15. Most prominently, the time to reach O-H dissociative CIs can be much longer than the ∼50 fs observed in the field-free case. This might be due to the longer time needed for a highly excited trajectory to reach the $S_1$ state in which the respective CI is accessible. In addition, the number of ring-openings observed is lower, and these can occur at higher energies compared to the field-free dynamics. This may result from the different paths on the PES the trajectories follow when excited to higher-lying excited states instead of $S_1$ and $S_2$ only, allowing the trajectories to reach higher-energy parts the $S_1$-$S_0$ intersection seam. For some trajectories, instead of one of the aforementioned relaxation pathways via CIs, return to the ground state is also observed in regions of the PES far away from CIs due to residual nonadiabatic couplings that invoke changes of the electronic state coefficients. This mechanism is particularly well possible if it is, as in the present case, preceded by the creation of an electronic superposition state due to interaction with the laser field. After return to the ground state, the majority of trajectories anew assumes the original cyclic structure, similar to the case of the field-free dynamics. In a few cases, reactions are observed, involving H atom abstraction in one instance, and 6 occurrences of a ring opening reaction, which may be also due to the larger internal energy of the photoexcited trajectories.

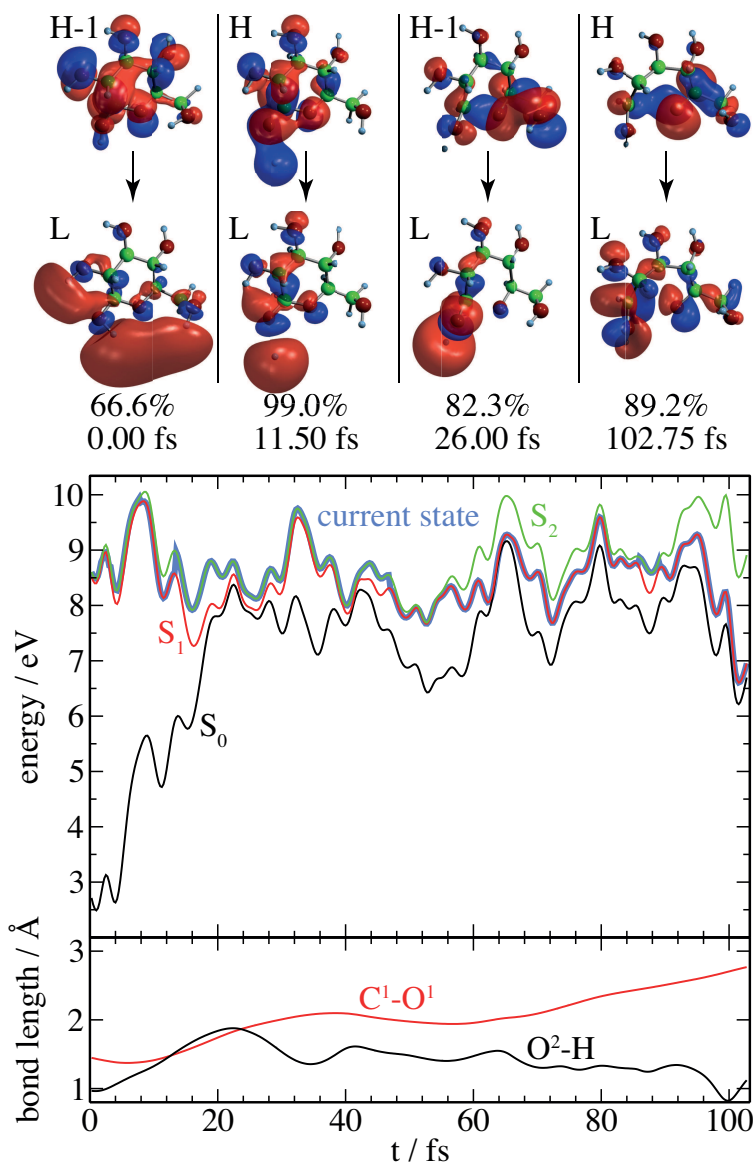

Figure 12: (Middle) Energies of the lowest 3 electronic states along a selected trajectory initiated in the $S_2$ state, where the $C^1$-$O^1$ ring-opening leads to a conical intersection. Energies are given relative to the optimized geometry of the ground state. (Top) Molecular orbitals showing the main excitation at distinct points in time. (Bottom) $O^2$-H and $C^1$-$O^1$ bond lengths along the trajectory.

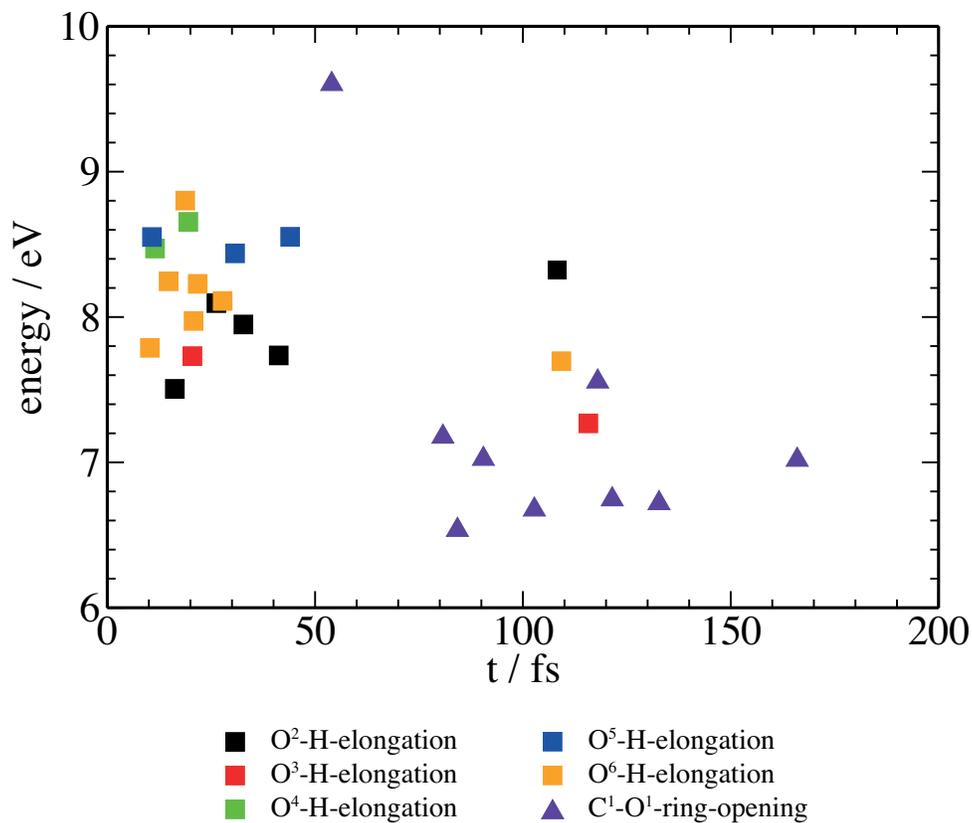

Figure 13: Ground state energies of the trajectories initiated in the $S_2$ state at the final time steps when the TDDFT energy gap between the $S_0$ and $S_1$ states closes. Energies are given relative to the optimized geometry of the ground state. The different types of CIs reached are classified according to the respective geometry changes as O-H elongation (squares) and as ring-opening (triangles). The CIs characterized by O-H elongation are further colour-coded according to which specific bond is affected.

## 4 Conclusions

The nonradiative relaxation of $\beta$-D-glucose from the lowest-lying excited electronic states takes place on a time scale of ∼ 100-200 fs. Return to the electronic ground state proceeds essentially quantitatively due to the presence of conical intersections (CIs) that can be easily accessed after photoexcitation. The geometries at which these CIs occur are consistent with those previously found by static calculations[14]: In most cases, elongation of O-H bonds takes place, while in some instances ring-opening due to dissociation of a C-O bond is observed. Although the O-H dissociative CIs tend to lie higher in energy, they are more easily reached due to the smaller geometric deformations necessary.

Since the UV absorption of glucose is characterized by the presence of many close-lying electronic states of similar oscillator strength, a direct simulation of the light-induced excitation process in the frame of the FISH method has been performed as well, using a field frequency corresponding to 6.43 eV photons. In this way, instead of a well defined single electronic state, a complex superposition is excited. Although this leads to a more involved electronic excitation and relaxation dynamics, the nuclear motion is qualitatively similar to the results of the field-free simulations, and the final return to the ground state occurs on the same time scale via the same types of CIs.

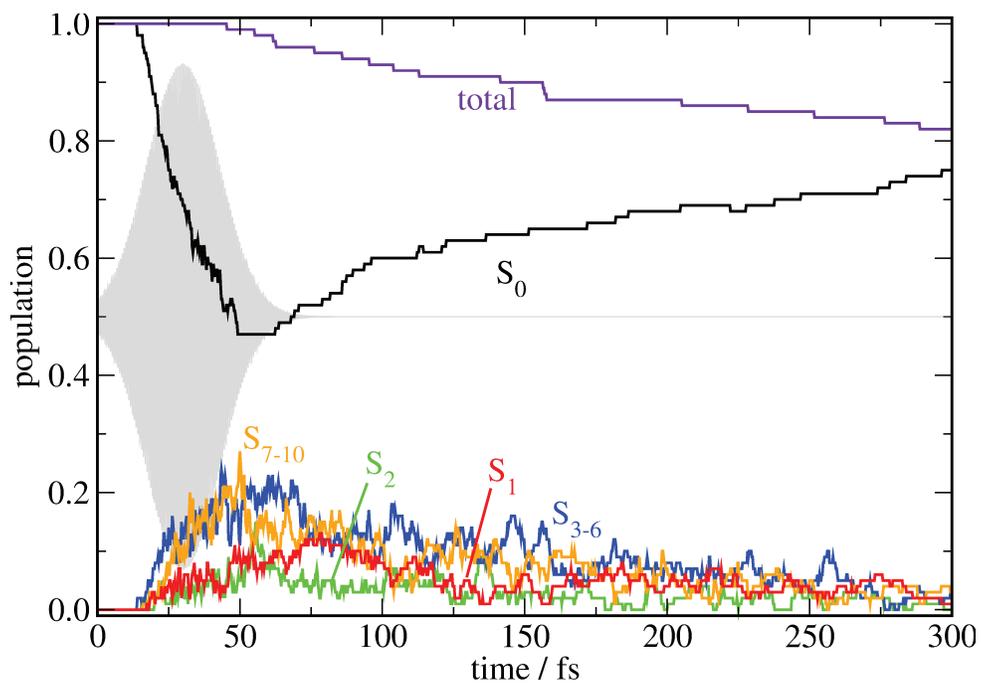

Figure 14: Time-dependent electronic state populations for the field-induced dynamics of glucose in the manifold of 10 electronic states. The excitation field is indicated in grey in the background. The decrease of the total population is due to such trajectories that resided in the higher excited states above $S_1$ when reaching the vicinity of an $S_0$-$S_1$ CI, where the propagation was stopped.

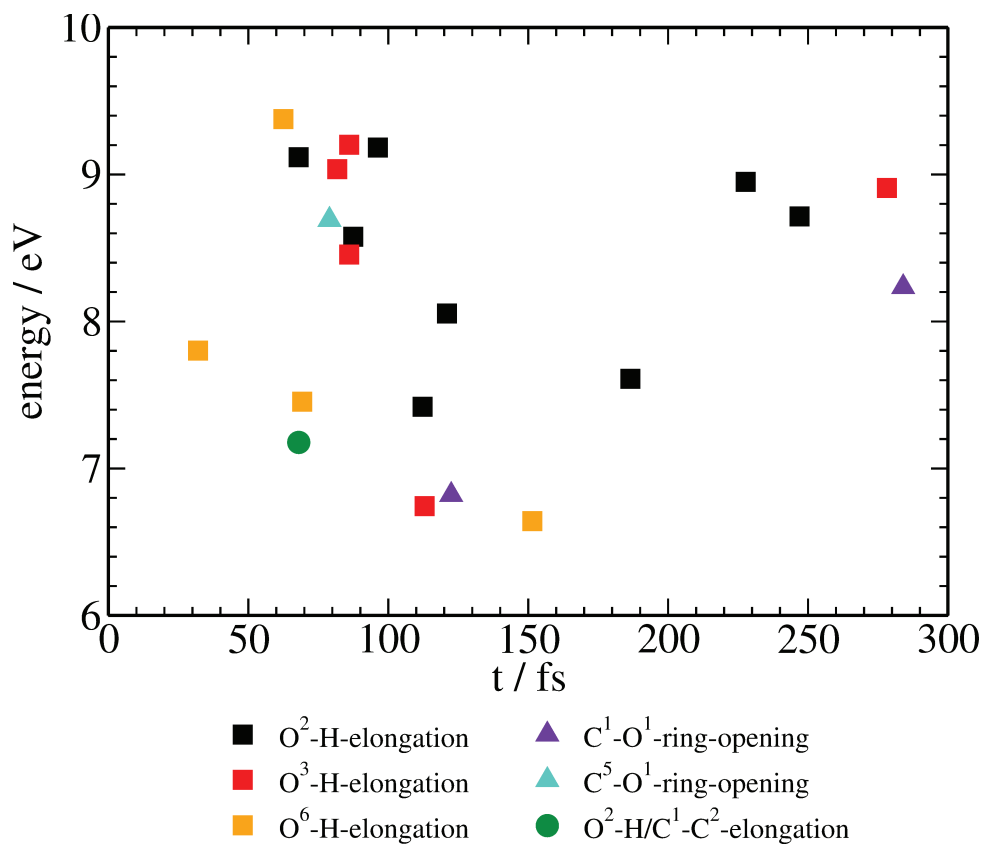

Figure 15: Ground state energies of the photoexcited trajectories at the final time steps when the TDDFT energy gap between the $S_0$ and $S_1$ states closes. Energies are given relative to the optimized geometry of the ground state. The different types of CIs reached are classified according to the respective geometry changes as O-H elongation (squares) and as ring-opening (triangles). The CIs characterized by O-H elongation are further colour-coded according to which specific bond is affected.

## Supporting Information

Additional computational details are provided: Cartesian coordinates of the optimized ground state structure of glucose, excitation energies and oscillator strengths of $S_1$ and $S_2$ obtained by TDDFT (B3LYP/6-31++G and 6-31++G**) and ADC(2)/aug-cc-pVDZ, electronic state populations and energies along selected nonadiabatic trajectories for TDDFT (B3LYP/6-31++G**) surface hopping dynamics, $S_0$ and $S_1$ energy scan along an O-H bond stretching coordinate for TDDFT (B3LYP/6-31++G, 6-31++G** and aug-cc-pVDZ) and ADC(2)/aug-cc-pVDZ.

## Acknowledgment

We wish to thank Dr. Deniz Tuna for drawing our attention to this topic during the Symposium on Theoretical Chemistry 2015 in Potsdam, Germany. Furthermore, we acknowledge financial support by the Deutsche Forschungsgemeinschaft, project FOR1809, as well as by the European Research Council, Consolidator Grant DYNAMO (Grant No. 646737).

## References

(1) Imberty, A.; Perez, S. Structure, conformation, and dynamics of bioactive oligosaccharides: Theoretical approaches and experimental validations. *Chem. Rev.* **2000**, *100*, 4567–4588.

(2) Simons, J. P.; Jockusch, R. A.; Çarçabal, P.; Hünig, I.; Kroemer, R. T.; Macleod, N. A.;

Snoek, L. C. Sugars in the gas phase. Spectroscopy, conformation, hydration, cooperativity and selectivity. *Int. Rev. Phys. Chem.* **2005**, *24*, 489–531.

(3) da Silva, C. O. Carbohydrates and quantum chemistry: how useful is this combination? *Theor. Chem. Acc.* **2006**, *116*, 137–147.

(4) Meinert, C.; Myrgorodska, I.; de Marcellus, P.; Buhse, T.; Nahon, L.; Hoffmann, S. V.; d'Hendecourt, L. L. S.; Meierhenrich, U. J. Ribose and related sugars from ultraviolet irradiation of interstellar ice analogs. *Science* **2016**, *352*, 208–212.

(5) Laurent, T. C.; Wertheim, E. M. Effect of ultraviolet light on absorption spectra of carbohydrates. *Acta Chem. Scand.* **1952**, *6*, 678–681.

(6) Laurent, T. C. Effect of ultraviolet light on alkaline solutions of glucose and certain other sugars. *J. Am. Chem. Soc.* **1956**, *78*, 1875–1877.

(7) Phillips, G. O.; Rickards, T. Photodegradation of carbohydrates. Part IV. Direct photolysis of D-glucose in aqueous solution. *J. Chem. Soc. (B)* **1969**, 455–461.

(8) Sobolewski, A. L.; Domcke, W. Photophysics of malonaldehyde: An ab initio study. *J. Phys. Chem. A* **1999**, *103*, 4494–4504.

(9) Sobolewski, A. L.; Domcke, W. Conical intersections induced by repulsive $^1\pi\sigma^*$ states in planar organic molecules: malonaldehyde, pyrrole and chlorobenzene as photochemical model systems. *Chem. Phys.* **2000**, *259*, 181–191.

(10) Sobolewski, A. L.; Domcke, W.; Dedonder-Lardeux, C.; Jouvet, C. Excited-state hydrogen detachment and hydrogen transfer driven by repulsive $^1\pi\sigma^*$ states: A new paradigm for nonradiative decay in aromatic biomolecules. *Phys. Chem. Chem. Phys.* **2002**, *4*, 1093–1100.

(11) Lucena Jr., J. R.; Ventura, E.; do Monte, S. A.; Araujo, R. C. M. U. Dissociation of ground and n$\sigma^*$ states of $CF_3Cl$ using multireference configuration interaction with sin-gles and doubles and with multireference average quadratic coupled cluster extensivity corrections. *J. Chem. Phys.* **2007**, *127*, 164320.

(12) Ashfold, M. N. R.; King, G. A.; Murdock, D.; Nix, M. G. D.; Oliver, T. A. A.; Sage, A. G. πσ* excited states in molecular photochemistry. *Phys. Chem. Chem. Phys.* **2010**, *12*, 1218–1238.

(13) Murdock, D.; Harris, S. J.; Luke, J.; Grubb, M. P.; Orr-Ewing, A. J.; Ashfold, M. N. R. Transient UV pump-IR probe investigation of heterocyclic ring-opening dynamics in the solution phase: the role played by nσ* states in the photoinduced reactions of thiophenone and furanone. *Phys. Chem. Chem. Phys.* **2014**, *16*, 21271–21279.

(14) Tuna, D.; Sobolewski, A. L.; Domcke, W. Electronically excited states and photochemical reaction mechanisms of β-glucose. *Phys. Chem. Chem. Phys.* **2014**, *16*, 38–47.

(15) Tuna, D.; Sobolewski, A. L.; Domcke, W. Conical-intersection topographies suggest that ribose exhibits enhanced UV photostability. *J. Phys. Chem. B* **2016**, *120*, 10729–10735.

(16) Roig, B.; Thomas, O. Rapid estimation of global sugars by UV photodegradation and UV spectrophotometry. *Anal. Chim. Acta* **2003**, *477*, 325–329.

(17) Sarazin, C.; Delaunay, N.; Costanza, C.; Eudes, V.; Mallet, J.-M.; Gareil, P. New avenue for mid-UV-range detection of underivatized carbohydrates and amino acids in capillary electrophoresis. *Anal. Chem.* **2011**, *83*, 7381–7387.

(18) Schmid, T.; Himmelsbach, M.; Oliver, J. D.; Gaborieau, M.; Castignolles, P.; Buchberger, W. Investigation of photochemical reactions of saccharides during direct ultraviolet absorbance detection in capillary electrophoresis. *J. Chromatogr. A* **2015**, *1388*, 259–266.

(19) Schmid, T.; Himmelsbach, M.; Buchberger, W. W. Investigation of photochemical reaction products of glucose formed during direct UV detection in CE. *Electrophoresis* **2016**, *37*, 947–953.

(20) Mitrić, R.; Petersen, J.; Bonačić-Koutecký, V. Laser-field-induced surface-hopping method for the simulation and control of ultrafast photodynamics. *Phys. Rev. A* **2009**, *79*, 053416.

(21) *TURBOMOLE, version V6.3*; Turbomole GmbH, Karlsruhe, Germany, 2011.

(22) Becke, A. D. Density-functional thermochemistry. 3. The role of exact exchange. *J. Chem. Phys.* **1993**, *98*, 5648–5652.

(23) Hehre, W. J.; Ditchfield, R.; Pople, J. A. Self-consistent molecular-orbital methods. 12. Further extensions of Gaussian-type basis sets for use in molecular-orbital studies of organic molecules. *J. Chem. Phys.* **1972**, *56*, 2257–2261.

(24) Clark, T.; Chandrasekhar, J.; Spitznagel, G. W.; von Ragué Schleyer, P. Efficient diffuse function-augmented basis sets for anion calculations. III. The 3-21+G basis set for first-row elements, Li - F. *J. Comput. Chem.* **1983**, *4*, 294–301.

(25) Hariharan, P. C.; Pople, J. A. Influence of polarization functions on molecular-orbital hydrogenation energies. *Theor. Chim. Acta* **1973**, *28*, 213–222.

(26) Dunning, T. H. Gaussian-basis sets for use in correlated molecular calculations. 1. The atoms boron through neon and hydrogen. *J. Chem. Phys.* **1989**, *90*, 1007–1023.

(27) Kendall, R. A.; Dunning, T. H.; Harrison, R. J. Electron-affinities of the 1st-row atoms revisited - Systematic basis-sets and wave-functions. *J. Chem. Phys.* **1992**, *96*, 6796–6806.

(28) Tully, J. C. Molecular dynamics with electronic transitions. *J. Chem. Phys.* **1990**, *93*, 1061–1071.

(29) Bauernschmitt, R.; Ahlrichs, R. Treatment of electronic excitations within the adiabatic approximation of time dependent density functional theory. *Chem. Phys. Lett.* **1996**, *256*, 454–464.

(30) Mitrić, R.; Werner, U.; Bonačić-Koutecký, V. Nonadiabatic dynamics and simulation of time resolved photoelectron spectra within time-dependent density functional theory: Ultrafast photoswitching in benzylideneaniline. *J. Chem. Phys.* **2008**, *129*, 164118.

(31) Mitrić, R.; Petersen, J.; Bonačić-Koutecký, V. In *Multistate nonadiabatic dynamics "on the fly" in complex systems and its control by laser fields*; Domcke, W., Yarkony, D. R., H., K., Eds.; in: Conical Intersections - Theory, Computation and Experiment, p. 497, Advanced Series in Physical Chemistry Vol. 17, Eds. W. Domcke, D. R. Yarkony, H. Köppel, World Scientific, Singapore, 2011.

(32) Lisinetskaya, P. G.; Mitrić, R. Simulation of laser-induced coupled electron-nuclear dynamics and time-resolved harmonic spectra in complex systems. *Phys. Rev. A* **2011**, *83*, 033408.

(33) Levine, B. G.; Ko, C.; Quenneville, J.; Martinez, T. J. Conical intersections and double excitations in time-dependent density functional theory. *Mol. Phys.* **2006**, *104*, 1039–1051.

(34) Schirmer, J. Beyond the random-phase approximation - A new approximation scheme for the polarization propagator. *Phys. Rev. A* **1982**, *26*, 2395–2416.

(35) Plasser, F.; Crespo-Otero, R.; Pederzoli, M.; Pittner, J.; Lischka, H.; Barbatti, M. Surface hopping dynamics with correlated single-reference methods: 9H-adenine as a case study. *J. Chem. Theor. Comput.* **2014**, *10*, 1395–1405.

(36) Dutta, A.; Sherill, C. D. Full configuration interaction potential energy curves for breaking bonds to hydrogen: An assessment of single-reference correlation methods. *J. Chem. Phys.* **2002**, *118*, 1610–1619.

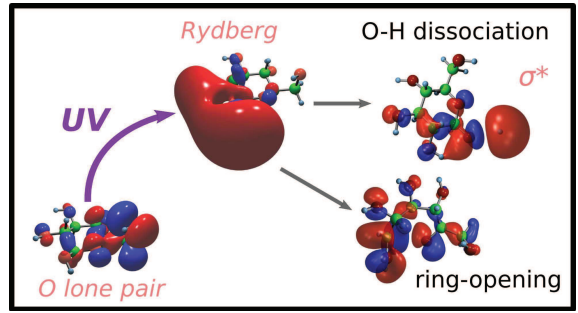

TOC Graphic

# Supporting Information:

# Ultrafast Photodynamics of Glucose

Jens Petersen,[*] Joachim O. Lindner, and Roland Mitrić[*]

*Julius-Maximilians-Universität Würzburg, Institut für physikalische und theoretische Chemie, Emil-Fischer-Str. 42, 97074 Würzburg, Germany*

E-mail: jens.petersen@uni-wuerzburg.de; roland.mitric@uni-wuerzburg.de

Phone: 0049 931 31 88832; 0049 931 31 85135

```
O    0.4425785   -1.4005781   -0.2078745
C   -0.9920354   -1.2947353   -0.4384342
C    1.2309564   -0.2297851   -0.6592670
O   -1.6202497   -2.4422834    0.1242136
C    2.6733396   -0.5646484   -0.2997864
H    1.1358423   -0.1216849   -1.7498507
C    0.6982570    1.0357827    0.0238594
O    1.3945078    2.1521014   -0.6011656
C   -0.8093450    1.1757813   -0.1700347
H    0.9294265    1.0012006    1.0950725
O   -1.2169021    2.3458541    0.5959337
H   -1.0262036    1.3481246   -1.2344534
C   -1.5428949   -0.0765062    0.2920252
O   -2.9511355    0.1387357   -0.0144094
H   -1.4051370   -0.227966     1.3696792
H   -3.4916842   -0.6079658    0.3135854
H   -2.1869412    2.4636568    0.5330705
H    1.0886865    2.9921970   -0.2020192
H    3.3136596    0.2933365   -0.5074477
O    2.8215000   -0.8432738    1.1220251
H    3.0050920   -1.4259765   -0.8940281
H    2.2191465   -1.5747827    1.3719849
H   -1.2862741   -3.2706160   -0.2748250
H   -1.1773819   -1.2307745   -1.5216727
```

Table S1: Cartesian coordinates of the optimized structure of $\beta$-D-glucose obtained by DFT (B3LYP, 6-31++G). Energy: $-686.6305948165\,E_H$

| State | TDDFT (B3LYP/6-31++G) | TDDFT (B3LYP/6-31++G**) | ADC(2) (aug-cc-pVDZ) |
|---|---|---|---|
| $S_1$ | 6.284 (0.0033) | 6.204 (0.0047) | 6.276 (0.0067) |
| $S_2$ | 6.410 (0.0042) | 6.462 (0.0037) | 6.617 (0.0084) |

Table S2: Lowest excitation energies (in eV) of $\beta$-D-glucose obtained by the different quantum chemical methods employed for surface hopping simulations. The oscillator strengths are given in parentheses.

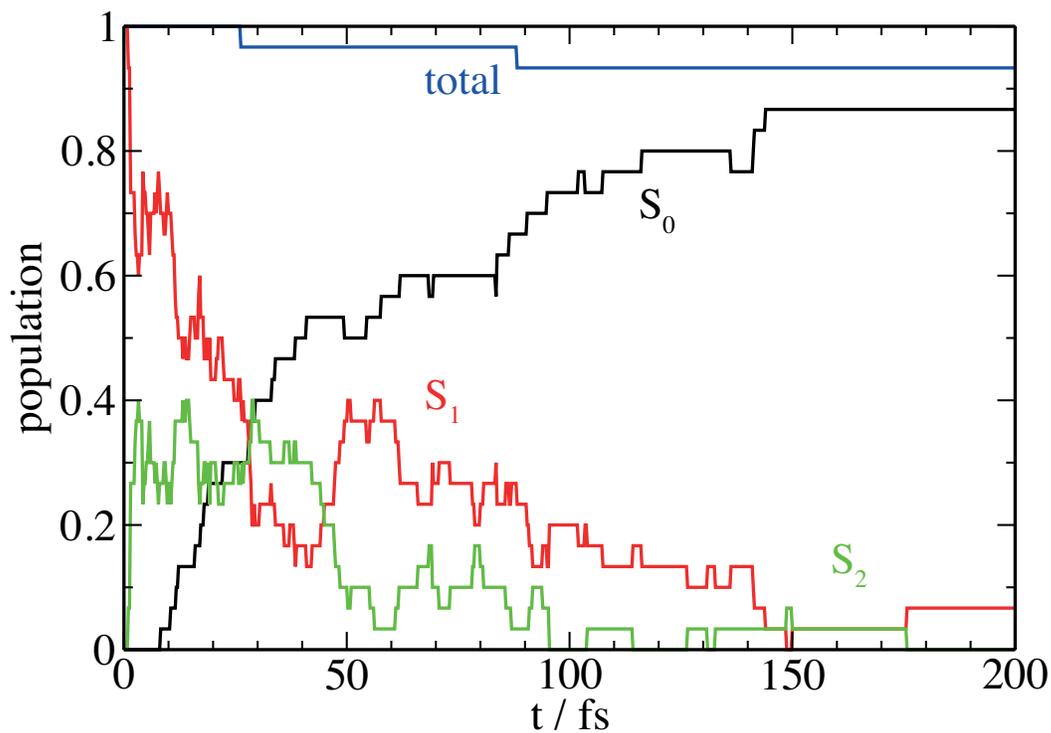

Fig. S1: Time-dependent electronic state populations for nonadiabatic surface-hopping dynamics of $\beta$-D-glucose starting in the $S_1$ state, obtained by TDDFT employing the B3LYP functional and 6-31++G** basis set. The ensemble comprised 30 trajectories whose initial conditions were sampled from a harmonic Wigner function at 250 K. The decrease of the total population is due to such trajectories that resided in the $S_2$ state when reaching the vicinity of an $S_0$-$S_1$ CI, where the propagation was stopped.

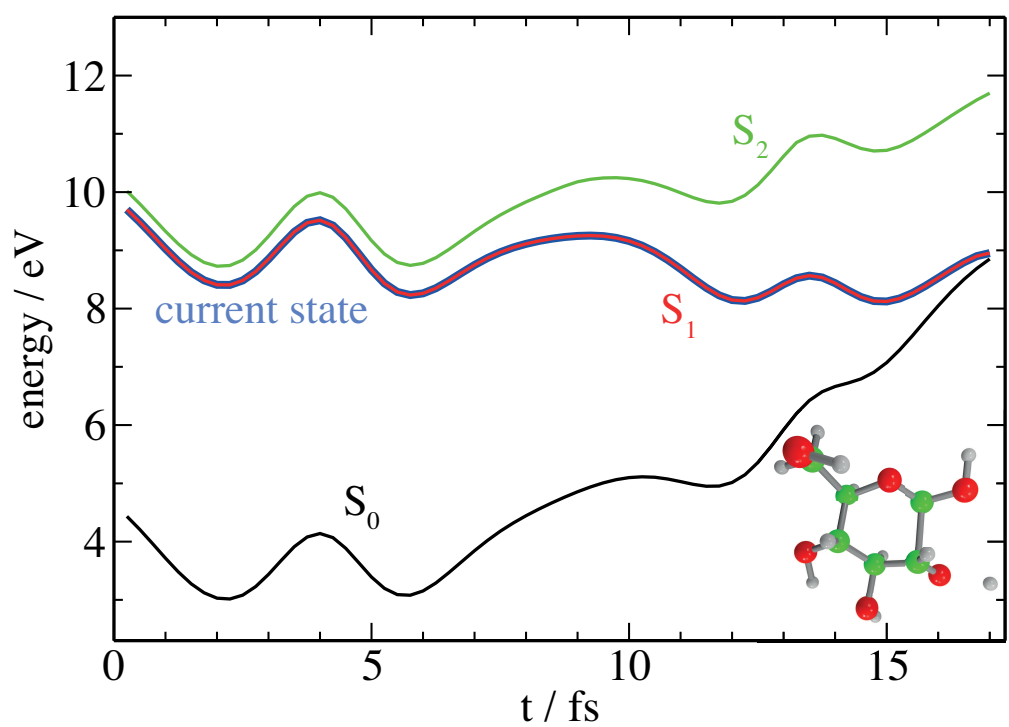

Fig. S2: Energies of the lowest 3 electronic states along a selected TDDFT (B3LYP/6-31++G**) trajectory initiated in the $S_1$ state, where the $O^2$-H-elongation leads to a conical intersection. Energies are given relative to the optimized geometry of the ground state.

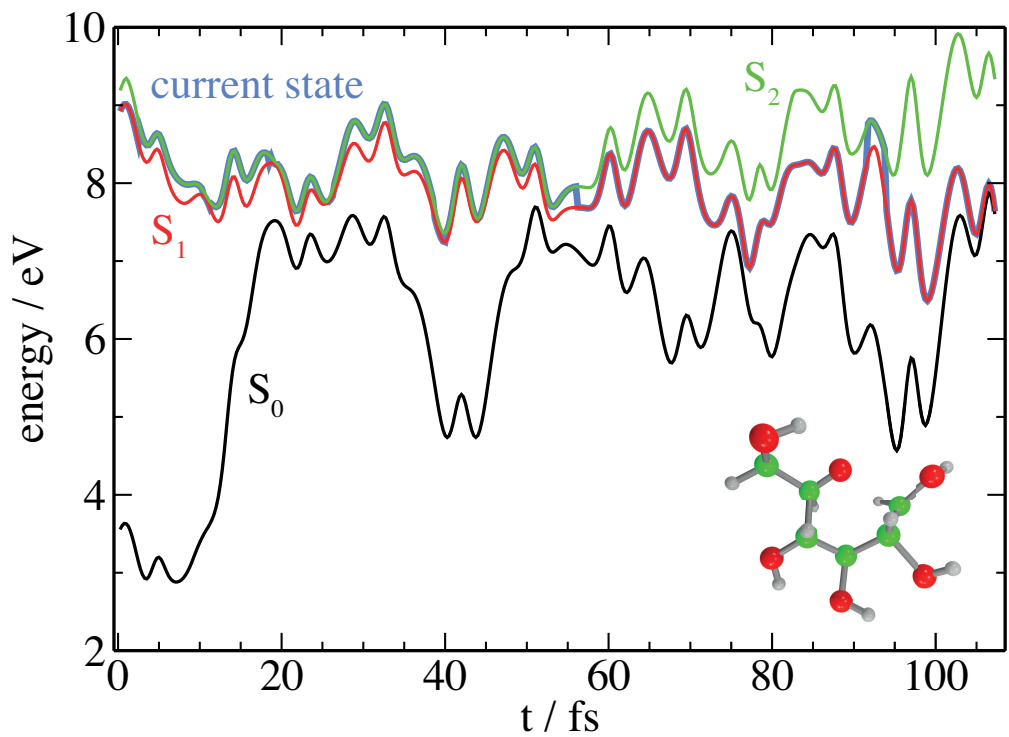

Fig. S3: Energies of the lowest 3 electronic states along a selected TDDFT (B3LYP/6-31++G**) trajectory initiated in the $S_1$ state, where the $C^1$-$O^1$ ring-opening leads to a conical intersection. Energies are given relative to the optimized geometry of the ground state.

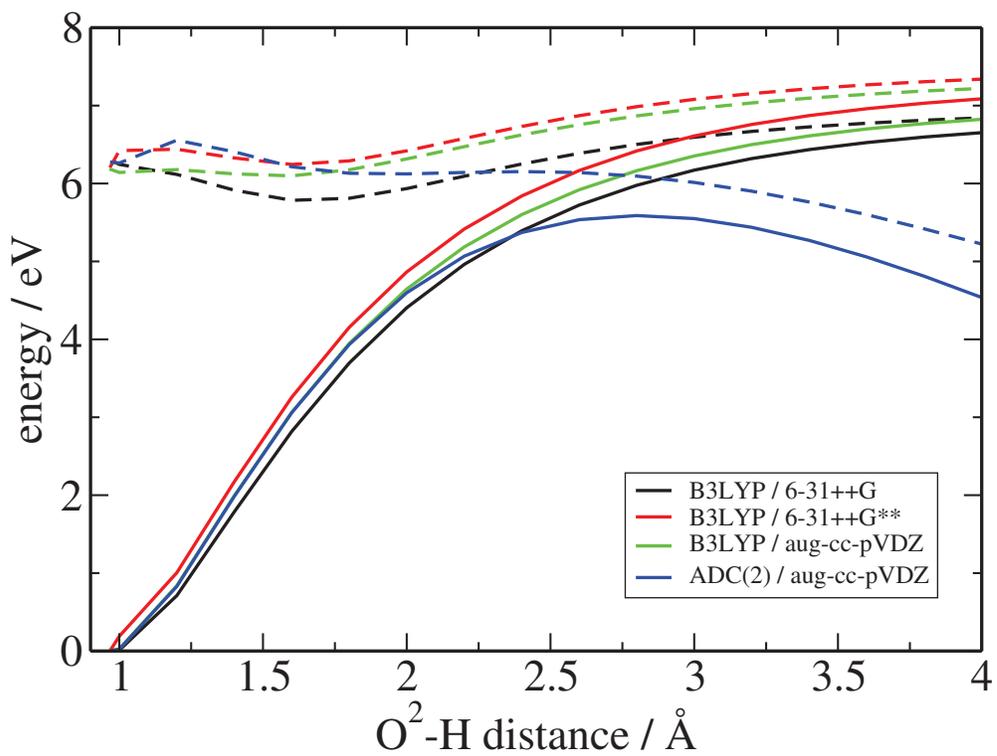

Fig. S4: Scan of the energies of the electronic ground state (full lines) and the first excited state (dashed lines) along the $O^2$-H bond stretching coordinate employing different quantum chemical methods.